\documentclass[conference]{IEEEtran}
\IEEEoverridecommandlockouts

\def\BibTeX{{\rm B\kern-.05em{\sc i\kern-.025em b}\kern-.08em
    T\kern-.1667em\lower.7ex\hbox{E}\kern-.125emX}}
\usepackage{booktabs}   
\usepackage{subcaption} 
\usepackage{cite}
\usepackage{amsmath,amssymb,amsfonts}
\usepackage{graphicx}
\usepackage{subcaption}
\usepackage{booktabs}
\usepackage{listings}
\usepackage{multirow}
\usepackage{pgfplots}
\pgfplotsset{compat=1.3}
\usepackage{algorithmic}
\usepackage{textcomp}
\usepackage{tabularx}
\usepackage{float}
\usepackage[latin1]{inputenc}
\usepackage{tikz}
\usepackage{xcolor}
\usepackage{url}
\usepackage{hyperref}
\usetikzlibrary{shapes,arrows}
\newcolumntype{L}[1]{>{\raggedright\let\newline\\\arraybackslash\hspace{0pt}}m{#1}}
\newcolumntype{C}[1]{>{\centering\let\newline\\\arraybackslash\hspace{0pt}}m{#1}}
\newcolumntype{R}[1]{>{\raggedleft\let\newline\\\arraybackslash\hspace{0pt}}m{#1}}

\renewcommand{\sl}{Skylake}
\newcommand{\mC}{\textit{MCompiler}}
\newcommand{\ml}{Machine Learning}

\begin{document}

\title{\mC{}: A Synergistic Compilation Framework
}

\author{\IEEEauthorblockN{Aniket Shivam}
\IEEEauthorblockA{\textit{Department of Computer Science} \\
\textit{University of California, Irvine}\\
Irvine, CA, USA \\
aniketsh@uci.edu}
\and
\IEEEauthorblockN{Alexandru Nicolau}
\IEEEauthorblockA{\textit{Department of Computer Science} \\
\textit{University of California, Irvine}\\
Irvine, CA, USA \\
nicolau@ics.uci.edu}
\and
\IEEEauthorblockN{Alexander V. Veidenbaum}
\IEEEauthorblockA{\textit{Department of Computer Science} \\
\textit{University of California, Irvine}\\
Irvine, CA, USA \\
alexv@ics.uci.edu}
}

\maketitle

\begin{abstract}
This paper presents a meta-compilation framework, the \mC{}.
The main idea is that different segments of a program can be compiled with different compilers/optimizers and combined into a single executable.
The \mC{} can be used in a number of ways.
It can generate an executable with higher performance than any individual compiler, because each compiler uses a specific, ordered set of optimization techniques and different profitability models and can, therefore, generate code significantly different from other compilers.
Alternatively, the \mC{} can be used by researchers and compiler developers to evaluate their compiler implementation and compare it to results from other available compilers/optimizers.

A code segment in this work is a loop nest, but other choices are possible.
This work also investigates the use of \ml{} to learn inherent characteristics of loop nests and then predict during compilation the most suited code optimizer for each loop nest in an application. 
This reduces the need for profiling applications as well as the compilation time.    

The results show that our framework improves the overall performance for applications over state-of-the-art compilers by a geometric mean of 1.96x for auto-vectorized code and 2.62x for auto-parallelized code.
Parallel applications with OpenMP directives are also improved by the \mC{}, with a geometric mean performance improvement of 1.04x (up to 1.74x).
The use of \ml{} prediction achieves performance very close to the profiling-based search for choosing the most suited code optimizer: within 4\% for auto-vectorized code and within 8\% for auto-parallelized code.
Finally, the \mC{} can be expanded to collect metrics other than performance to be used in optimization process. The example presented is collecting energy data.
\end{abstract}


\section{Introduction} \label{sec:intro}
An important compiler task is optimizing applications for better performance on a target architecture.
Optimizing loop nests, in particular,  contributes significantly towards achieving better performance.
State-of-the-art architectures have multiple cores on a chip, where each core has Single Instruction Multiple Data (SIMD), or vector, capabilities.
These architectural features provide opportunities for a compiler to expose parallelism in applications on multiple levels.
The code optimization techniques to \textit{auto-vectorize} the loop nests~\cite{Padua86,Allen87,Wolfe95}, so as to generate SIMD instructions, require careful analysis of data dependences, memory access patterns, etc. 
Several auto-parallelization techniques~\cite{Padua80,Li92,Lim98,Lim99,Lim01,Bondhugula08b,Darte12} and directive based parallel programming models, such as OpenMP~\cite{openmp5.0}, have been developed to take advantage of multiple cores. 
In fact, most auto-parallelization implementations in modern compilers, which take serial code as input, generate OpenMP code.  

Key loop transformation techniques~\cite{Banerjee93,Wolfe95,Kennedy01} include Distribution, Fusion, Interchange, Skewing, Tiling and Unrolling.
Code optimizers search for an optimal semantic-preserving sequence of transformations to generate a better performing code, either serial or parallel.
But evaluating if a sequence of transformations is optimal is complex and the search for the best sequence of transformations and their profitability is guided by heuristics and/or approximate analytical models. 
Thus, a code optimizer may end up with a sub-optimal result and different code optimizers may, for the same source code segment, generate code with significant performance differences on the same architecture.
Some major challenges in developing the heuristics and profitability models is predicting the behavior of a multi-core processor which has complex pipelines, multiple functional units, complex memory hierarchy, hardware data prefetching, etc. 
Parallelization of loop nests involves further challenges for the code optimizers, since communication costs based on the temporal and spatial data locality among iterations have an impact on the overall performance.
Evaluation studies~\cite{Moseley09,Tournavitis09,Maleki11,Gong18} have shown that state-of-the-art code optimizers may miss out on opportunities to auto-vectorize and auto-parallelize the loop nests for modern architectures. 
For optimizing applications written in \texttt{C}, there are several compilers and domain specific loop optimizers that perform auto-vectorization and, in some cases, auto-parallelization of code.
From a given code optimizer's point of view, the sequence it used is the best but there is no way of knowing how close it gets to optimal performance or if there is any headroom for improvement.

This paper presents a compiler framework, \mC{}, that allows each loop nest to be optimized by the best optimizer available to it. The \mC{} incorporates code optimizers from Intel's C compiler, PGI's C compiler, GNU GCC, LLVM Clang. 
In addition to these, two Polyhedral Model based loop optimizers, Polly~\cite{Grosser12,Polly} and Pluto~\cite{Bondhugula08a,Pluto} are used. 
The \mC{} identifies loop nests from \texttt{C} applications, optimizes the loop nests using different code optimizers, profiles each optimized code version as part of the applications, and links the best performing codes to generate the complete application binary. 
The best loop nest code selection allows the \mC{} to produce higher-performing code than the best of the compilers in the framework. 
The framework allows for easy integration of newer versions and newer configurations of the available code optimizers and also allows the addition of new code optimizers.

The framework can be used to optimize applications, first, for serial execution with auto-vectorization of loop nests.
This optimizes loop nests for SIMD or vector code generation, in addition to optimizing loop nests for data locality, memory hierarchy, etc.
Second, our framework can also target multi-core processors, by taking serial loop nest codes as input and auto-parallelizing those loop nests using the available code optimizers to generate multi-threaded code. 
Auto-parallelized code is also optimized for SIMD execution over each thread.
In this case, the original loop nests are transformed such that loop iterations can be reordered and scheduled for parallel execution across the multiple cores.
Third, our framework can target OpenMP applications, i.e., applications with OpenMP directives inserted across sections of the code meant for parallel execution.
The  performance evaluation of the \mC{} shows that our framework indeed achieves an overall geometric mean speedup of 1.96x for serial code, 2.62x for auto-parallelized code and 1.036x (up to 1.74x) for OpenMP code over Intel C Compiler for applications from various benchmark suites.

The framework extracts loop nests from the applications' source files into separate source files as a function, together with any additional information needed. It then replaces loop nests with a function call in the original source files.
This allows for separate code optimizers to focus on just the loop nests and also allows the framework to insert the best performing code, i.e., linking object files to generate the executable. 
To evaluate the speedup potential, the framework initially optimizes each extracted loop nest with all available code optimization \textit{candidates}.
The performance of each optimized loop nest is measured as part of the complete application and allows for selecting the best performing code for each loop nest.
This step is highly time consuming, it was used to establish that the framework can indeed improve performance. 
The final step links the selected object files 
 generating the executable for the complete application.

The second goal of this paper is to use \ml{} (ML) models to predict the most suited code optimizer for a given loop nest.
This can eliminate the profiling in the framework and should reduce compilation time.
However, this can lead to a potential performance loss compared to profiling due prediction errors.
The \ml{} model or \textit{classifier} can predict a different optimizer than the profiling-based best code optimizer.

The input or \textit{features} to our \ml{} model are hardware performance counters collected from profiling a serial (-O1) version of a loop nest.
Embedding Machine Learning models in compilers is continuously being explored by the research community~\cite{Monsifrot02,Stephenson05,Cavazos07,Tournavitis09,Wang09,Fursin11,Stock12,Cammarota13,Watkinson17,Ashouri17,Shivam18}.
Most of the previous work used \ml{} in the domain of auto-vectorization, phase-ordering and parallelism runtime settings.
This work applies Machine Learning on a coarser level, in order to predict the most suited code optimizer - for serial as well as parallel code.

Previous studies have shown that hardware performance counters can successfully capture the characteristic behavior of a loop nest.
Machine Learning models in those studies either used a mix of static features (collected from source code at compile time) and dynamic features (collected from profiling)~\cite{Tournavitis09,Wang09}, or exclusively use dynamic features~\cite{Cavazos07,Watkinson17,Ashouri17,Shivam18}.
The approach used in this paper belongs to the latter class and exclusively uses hardware performance counters collected for a loop nest.
The framework makes the predictions for the most suited code optimizer using the incorporated, trained ML model as it compiles each loop nest.
This replaces the expensive profiling for every optimizer with a single profile of each nest to collect hardware performance counters. They are then used as features to make the prediction from the trained ML classifier.
The functionality to predict the most suited code optimizer for loop nests can also be embedded in a Just-In-Time (JIT) compiler, but this is part of future work.
It would use run-time profiling, for instance of a subset of iterations, to collect hardware performance counters and then use them to make predictions.

The evaluation of the \mC{} with \ml{} predictions shows that the performance of applications is within 4\% for auto-vectorized code and within 8\% for auto-parallelized code compared to the profiling-based search for the most suited code optimizer.
We exclude OpenMP applications from ML predictions because in this case it is the user-inserted directives that make most of the difference to the performance rather than the inherent characteristics of loop nests.
Hence, this problem is not suitable for such predictions.

The paper also highlights the usability of the \mC{} framework for compiler researchers to evaluate their optimization techniques or improvements against other available code optimizers and compilers.
To compile an application using the \mC{} framework, the user needs a few modifications to the build configurations, similar to what is required to add any other compiler. But no modifications to the source code of the applications are required.

The paper also shows how the \mC{} framework can be extended with additional metrics to allow users to gain insight into energy consumption on target architecture.
This highlights the potential for the expansion of the \mC{} framework. 
%
\begin{figure*}[ht]
  \centering
  \includegraphics[width=0.8\paperwidth]{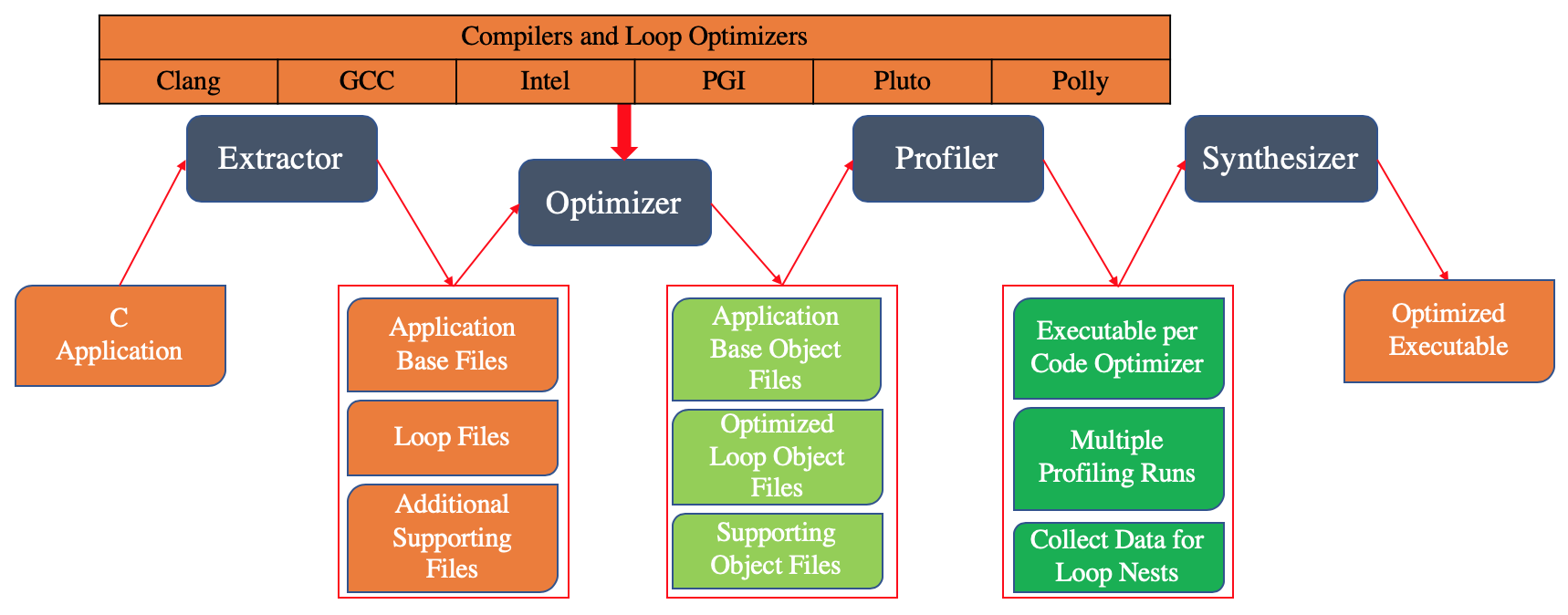}
  \caption{\mC{} Framework}
  \label{fig:framework} 
\end{figure*}

Overall, this paper makes the following contributions:
\begin{itemize}
\item It presents a meta-compilation framework that improves performance for \texttt{C} applications for serial as well as parallel execution, including OpenMP applications.
\item It demonstrates that prediction for the most suited code optimizer (serial as well as parallel) for a loop nest can be accurately made using \ml{} classifiers.
\item It presents the \mC{} framework extension for reporting  energy consumption per loop nest.
\item The framework will be open sourced for researchers and compiler developers to analyze and compare their code optimization techniques.
\end{itemize}

The rest of the paper is organized as follows. Section \ref{sec:implementation} describes the \mC{} framework and the methodology for choosing the most suited code optimizer for a loop nest using profiling-based search as well as using ML-based prediction. Section \ref{sec:result} describes the evaluation methodology and analyzes the experimental results. Section \ref{sec:priorart} discusses related work. We conclude the paper with Section \ref{sec:summary}.

\section{Framework Design and Implementation} \label{sec:implementation}
This section describes the overall architecture of the \mC{} framework and the technical details about the individual phases of the framework.
The specifics of the architecture for incorporating \ml{} predictions is discussed later in the section.
\begin{table*}[!ht]
\centering
\begin{tabular}{|C{0.15\paperwidth}|C{0.06\paperwidth}|C{0.35\paperwidth}|C{0.12\paperwidth}|}
\hline
 \textbf{Code Optimizer} & \textbf{Version} & \textbf{Flags (Auto-Parallelization flags)} & \textbf{Auto-Parallelization} \\ \hline
 clang (LLVM)                   & 7.0.0   & -Ofast -march=native                    & No  \\ \hline
 gcc (GNU)                      & 5.4.0   & -Ofast -march=native                    & No  \\ \hline
 icc (Intel)                    & 18.0.0  & -Ofast -xHost (-parallel)               & Yes \\ \hline
 pgcc (PGI)                     & 18.10   & -fast -tp=skylake -Mllvm (-Mconcur)     & Yes \\ \hline
 PLuTo (source-to-source) + icc & 0.11.4  & \texttt{--}tile (\texttt{--}parallel)   & Yes \\ \hline
 polly                          & 7.0.0   & -O3 -march=native -polly  
 											-polly-tiling
                                            -polly-vectorizer=stripmine 
                                            (-polly-parallel)                       & Yes \\ \hline
\end{tabular}
\caption{Candidate Code Optimizer and their flags}
\label{table:compilerflags}
\end{table*}
\subsection{Overall Framework Architecture}
Figure \ref{fig:framework} shows the structure of the \mC{} framework.
The first phase is the \textit{Loop Extraction} from \texttt{C} applications.
Since loop nests dominate the performance in various applications, the \textit{Extractor} parse the source files to find \texttt{for} loop nests, extract those loop nests as \texttt{functions} into separate independently compilable files and replace the loop nests with the corresponding function call in the \textit{base} source file.
Base files are similar to the original source files but with loop nests replaced with function calls.
Whereas loop files are newly generated files which define the function containing the loop body and supporting components to make them compile successfully.
This Extractor is inspired by the loop extractor described in the work by Chen et. al.~\cite{Chen17} to encapsulate loop nests into standalone executables.

The second phase is the \textit{Optimization} phase. 
The \textit{Optimizer} compiles each loop file with the available code optimizers.
Also, it compiles the base files and additional \mC{} files, i.e., files added to support the functioning of the framework.
For source-to-source code optimizers, a \textit{default compiler} is used to compile optimized loop files, the base files and additional files.

The third phase is the \textit{Profile} phase, where the application is profiled for execution times of the extracted loop nests.
Executables generated for each code optimizer is executed and reported execution times for the loop nests are collected. 

The final phase is the \textit{Synthesis} phase.
Here, for each extracted loop nest, the collected loop execution times are compared from every code optimizer and a code optimizer that produced the best performing code is selected, i.e., the optimized code that completes the execution of the loop body in the shortest time.
Finally, the default compiler is used to link the object files from the selected code optimizer for every loop file, plus the object files generated by the default compiler for the base files.
This step also requires linking libraries that code optimizers may have used or taken support of for generating code for the loop files.

For large applications, if \texttt{-c} flag is provided, i.e., compile to object files only, then just the Extractor and the Optimizer are enabled.
In such cases, the Profiler and the Synthesizer are enabled only at link-time.
\mC{} framework handles flags for macro definitions, paths to header files and libraries for linking, etc. similar to other compiler.

\subsection{Loop Extraction Phase}
The loop extractor works in three phases and is implemented using \texttt{ROSE}, a source-to-source compiler infrastructure~\cite{Quinlan00}.
First, the extractor traverses the abstract syntax tree (AST) and locates the \texttt{for} loop nests that are eligible for extraction.
Second, the extractor creates a new file for this loop, adds necessary headers and macro definitions in the loop file, and also add \texttt{extern} declaration for global variables and global functions, as well as, for functions called in the scope of the loop body.
Encloses the loop body in a function definition with parameters being the variables and pointers to the data structures required by the loop body in order to compile and run correctly.
Third, in the base file's AST replace the loop body with a function call (with required arguments) and add an \texttt{extern} declaration to this function.
Finally, generate the modified base source file and the new loop files.

While traversing the AST for eligible loop nests, the extractor skips loop nests with irregular control flow that hinders extraction, i.e., contains \texttt{return} and \texttt{goto} statements.
Also, it skips loop nests with calls to static functions and static variables since those properties hinder their usage in the new loop files.

The extractor generates two similar versions of the loop files, where one version contains extra code around the loop body to collect profile information about the loop nests.
The version with the profile code is used during the Profile phase and is responsible for generating information regarding the execution time of the loop nests.
The other does not contain any profile code and is used while generating the final executable for the applications. 

\subsubsection{Function Definition enclosing the Loop Nests}
The extractor generate the lists of variables, with their data types, used inside the scope of the loop body.
All primitive data types (\texttt{int}, \texttt{float}, etc.) are passed by reference, as well as the user-defined types such as arrays, \texttt{structs} and \texttt{typedefs}.
The extractor also does an optimization to maintain properties of the loop from the point of view of the code optimizers.
This optimization copy the function parameters of primitive types (passed by reference) into local variables (with same names as original variables) before the loop body and correspondingly copy the local variables into the function parameters at the end of the loop body.
This optimization prevents any change to loop body and is also critical to performance since usage of pointers can prevent some code optimizations.

For loop nests with OpenMP directives, the extractor moves the directives with loop body and sanitizes the clauses of variables that are not present in the scope of the loop nest.
For OpenMP \texttt{for} loops that are enclosed in a \texttt{omp parallel} region, extracting the loop body with \texttt{omp for} directive doesn't change the behavior of the program.
One drawback of extracting OpenMP \texttt{for} loops that are enclosed in a \texttt{parallel} region in such manner is that in the presence of \texttt{threadprivate} variables, synthesizer encounters a link-time error because compilers may generate different symbols for the same \texttt{threadprivate} variable.
\begin{figure*}[ht]
  \centering
  \includegraphics[width=0.8\paperwidth]{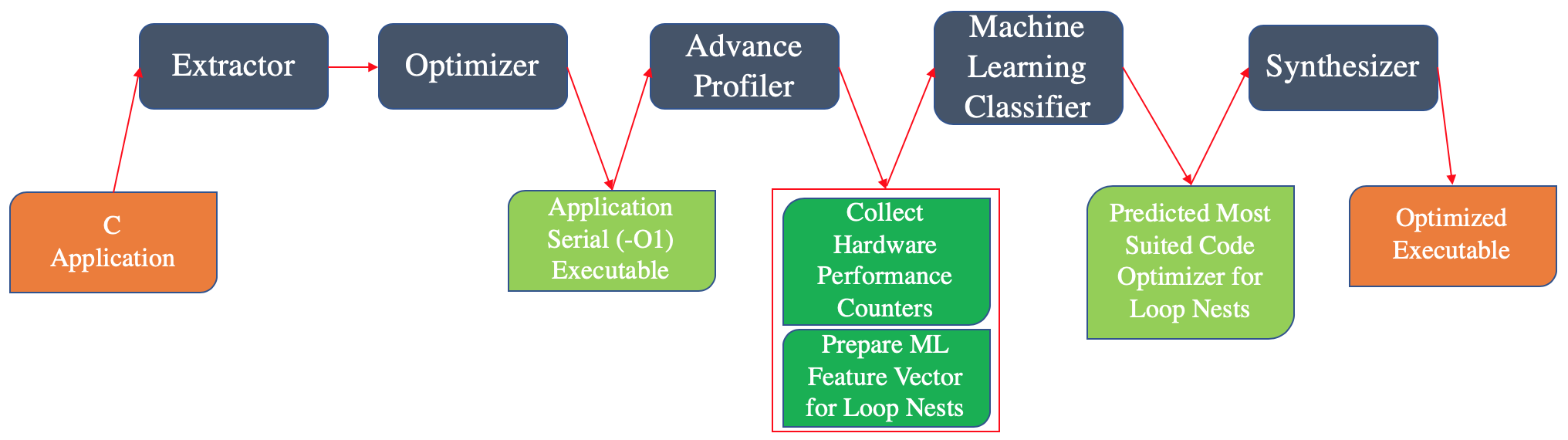}
  \caption{\mC{} Framework with \ml{} Predictions}
  \label{fig:ml_framework} 
\end{figure*}
\subsection{Optimization Phase}
The framework uses six candidate code optimizers: Intel's \texttt{icc}, PGI's \texttt{pgcc}, GNU's \texttt{gcc}, LLVM \texttt{clang}, LLVM based polyhedral loop optimizer \texttt{Polly} and source-to-source polyhedral loop optimizer \texttt{Pluto}.
We chose \texttt{icc} as the default compiler because its performance is, on average, the best of the compilers included. It is also used to compile source files generated by source-to-source loop optimizer, i.e., Pluto. 
Table \ref{table:compilerflags} shows the flags used for optimizing loop nests for serial execution and parallel execution.
These flags also include target architecture specific flags to enable optimizations that can generate better performing code on the specific architecture.
For OpenMP applications, flags from serial configuration are used in addition to the OpenMP flags.

The optimizer can compile loop files and base files in parallel.
This is similar to \texttt{-j} option of Makefiles, but here all candidate code optimizers are invoked in parallel to compile the source files.
This reduces the overall compilation time for the \mC{} framework.
The optimizer generates multiple executables of the application (with profile code) where each executable is completely compiled and linked by a candidate code optimizer. 

\subsection{Profile Phase}
The profiler executes the executables generated by the code optimizers one-by-one and performs multiple runs for stable data, if requested.
Profiler at the end of each execution collects the profiled information for each of the loop nests and forwards it to the Synthesizer.
For applications that need input through command line, the profiler runs the application with the input given to the \mC{} framework using a \texttt{--input} flag.

\subsection{Synthesis Phase}
The synthesizer compares the collected profile information, i.e., the execution times for each loop nests from different code optimizers and chooses the code optimizer that performed the best as the most suited code optimizer.
For loop nests with no profile information, i.e., the code that was not executed during profiling, the default compiler is chosen as the most suited code optimizer.
The synthesizer then generates the final executable that contains no profile code.
For OpenMP application, the synthesizer links OpenMP runtime libraries that are used by different compilers, e.g., \texttt{icc} and \texttt{clang} use compatible OpenMP runtime libraries whereas \texttt{gcc} doesn't.
Static libraries specific to compilers are also linked to successfully generate the final executable.

\subsection{Framework Architecture for \ml{} Predictions}
The framework for choosing the most suited code optimizer for the loop nests using the ML predictions is shown in figure \ref{fig:ml_framework}.
The ML predictions are used to predict the most suited code optimizer for both serial, auto-vectorized code as well as auto-parallelized code.
The input to the ML Classifier for making the predictions are the hardware performance counters for the loop nests.
This strategy for collecting hardware performance counters for the loop nests and using them to predict the most suited code optimizer is inspired from the work of Shivam et. al.~\cite{Shivam18}.
The architecture of the \mC{} framework is modified in the following ways to predict the most suited code optimizer for the loop nests.

First, the Optimizer now generates an additional executable that is compiled by the default compiler for serial execution with -O1 optimization level.
Second, the Profiler is replaced by the Advance Profiler for making ML predictions.
Advance Profiler executes the serial (-O1) execution and collect hardware performance counters for the loop nest.
If the loop nest is not executed or the hardware performance counters are not present (happens for loop nests with very few computations), the default compiler is chosen by the Synthesizer.

Next, the collected hardware performance counters for each loop nest are transformed into the feature vector, i.e., the input to the ML classifier.
Third, the ML classifier makes the prediction for the most suited code optimizer for a loop nest based on the feature vector.
The ML classifier is a trained ML model.
There are two separately trained ML models, one for serial code predictions while the other is for parallel code predictions.
Finally, these predictions from the ML classifier are forwarded to the Synthesizer, which uses the code optimizer from the prediction to link the correct optimized loop object files and generate the final executable for the application.

The \mC{} driver invokes the ML prediction part of the framework over the original \mC{} flow with profiling-based search if the \texttt{--predict} flag is provided.
The ML models are trained and incorporated in the \mC{} framework using OpenCV's \ml{} module~\cite{opencv4.0}.

\subsubsection{Collecting Hardware Performance Counters for the Loop Nests}
The features, i.e., the hardware performance counters used for the Machine Learning models are collected by profiling loop nests using Intel's VTune Amplifier.   
We use generated code from Intel compiler to generate the executable that is then used for profiling.
All the loop optimizations are disabled during this compilation by using the -O1 flag.
In addition to that, the optimization that are responsible for vector code generation and parallel code generation are disabled too.
The profiling information, therefore, provides an insight into the characteristics of the loop nests while eliminating the influence of compiler transformations and behavioral changes incurred from special architectural features of the underlying architecture.
The performance counters that are collected include, but not limited to, instruction-based (instruction types and counts) counters, CPU clock cycles-based (including stalls) counters, memory-based (D-TLB, L1 cache, L2 cache, L3 cache) counters.

Once the hardware performance counters are collected for the loop nests, we skip dynamic instruction count as a feature and normalize the rest of the hardware performance counters in terms of \textit{per kilo instructions} (PKI).
Based on our analysis, this allows the Machine Learning models to learn about the inherent characteristics of the loop nests and not bias them towards characteristics such as loop trip count.
\begin{figure*}[ht]
  \centering
  \includegraphics[width=0.8\paperwidth]{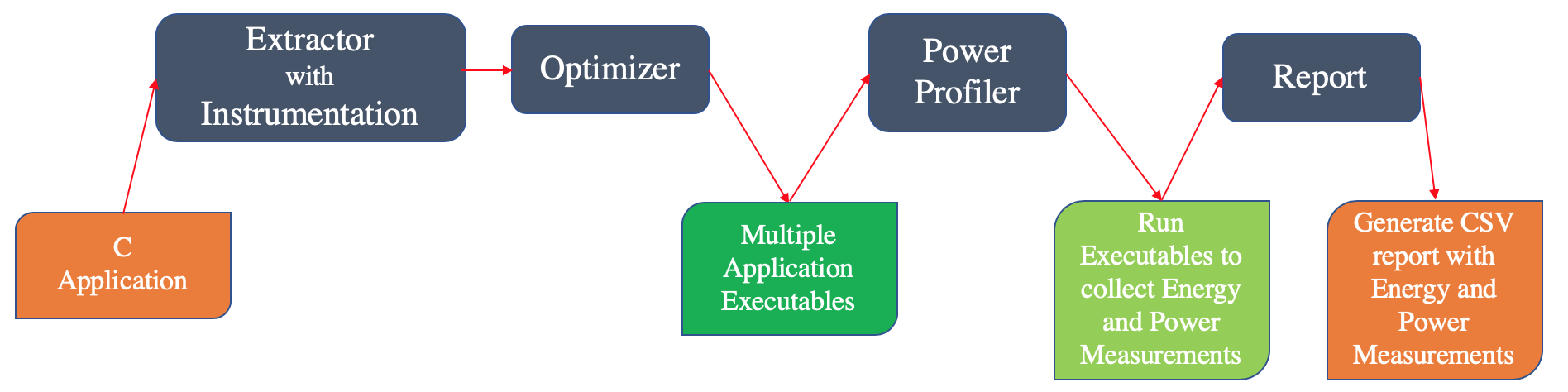}
  \caption{\mC{} Framework for collecting Energy and Power Measurements}
  \label{fig:power_framework} 
\end{figure*}
\subsubsection{Random Decision Forest Classifier (RF)}
We chose Random Decision Forest~\cite{Ho95} as our classification algorithm to predict the most suited code optimizer for the loop nests. RF is a learning algorithm that builds on the principles behind Decision Trees.
Generally, the Decision Tree algorithm, learns from training data by building a structured and hierarchical representation of the correlation between features and classes.
Features represent the nodes in the trees and classes are leaves at the deepest level.
An optimal Decision Tree would perfectly and accurately divide the data among the target classes.
However, finding an optimal tree is an NP-Complete problem, therefore we have to rely on heuristics such as greedy search.

%
Decision Trees suffer from several issues.
The main one being a tendency towards overfitting, that is, the tree loses generalization the deeper the tree goes, modeling the trend for training data but be inaccurate for new instances.
RF provides a better solution for overfitting and classification bias by adding two stochastic steps to the Decision Tree Algorithm.
From the training dataset, RF creates a bootstrapped subset by stochastically choosing the instances or features (with repeats allowed) that will be used for building the decision trees.
This is called Bootstrap Aggregating and the instances inside the created subset are called In-Bag instances.
After creating the Bagged dataset, an arbitrary number of decision trees are built using subsets of randomly chosen features.
The depth of the trees is limited by the number of features allowed (according to a predetermined threshold) and by another arbitrary number.
Each decision tree accuracy is evaluated using the remaining (out-of-bag) instances that weren't part of the Decision Tree building phase.

Classification is achieved through a voting algorithm, where a target value is generated by each Random Tree, the one with the highest number of trees will be the class assigned to the new instance.

Since Random Forest is constantly evaluating the performance of the subsets of features, we can easily detect the ones that were used in the better performing trees. 
Therefore, we require little to no feature filtering before running the algorithm.
This is useful when approaching a new problem where the correlation between the input features and the output class is not entirely known.
\begin{figure}[ht]
  \centering
  \includegraphics[height=9cm,width=0.95\columnwidth]{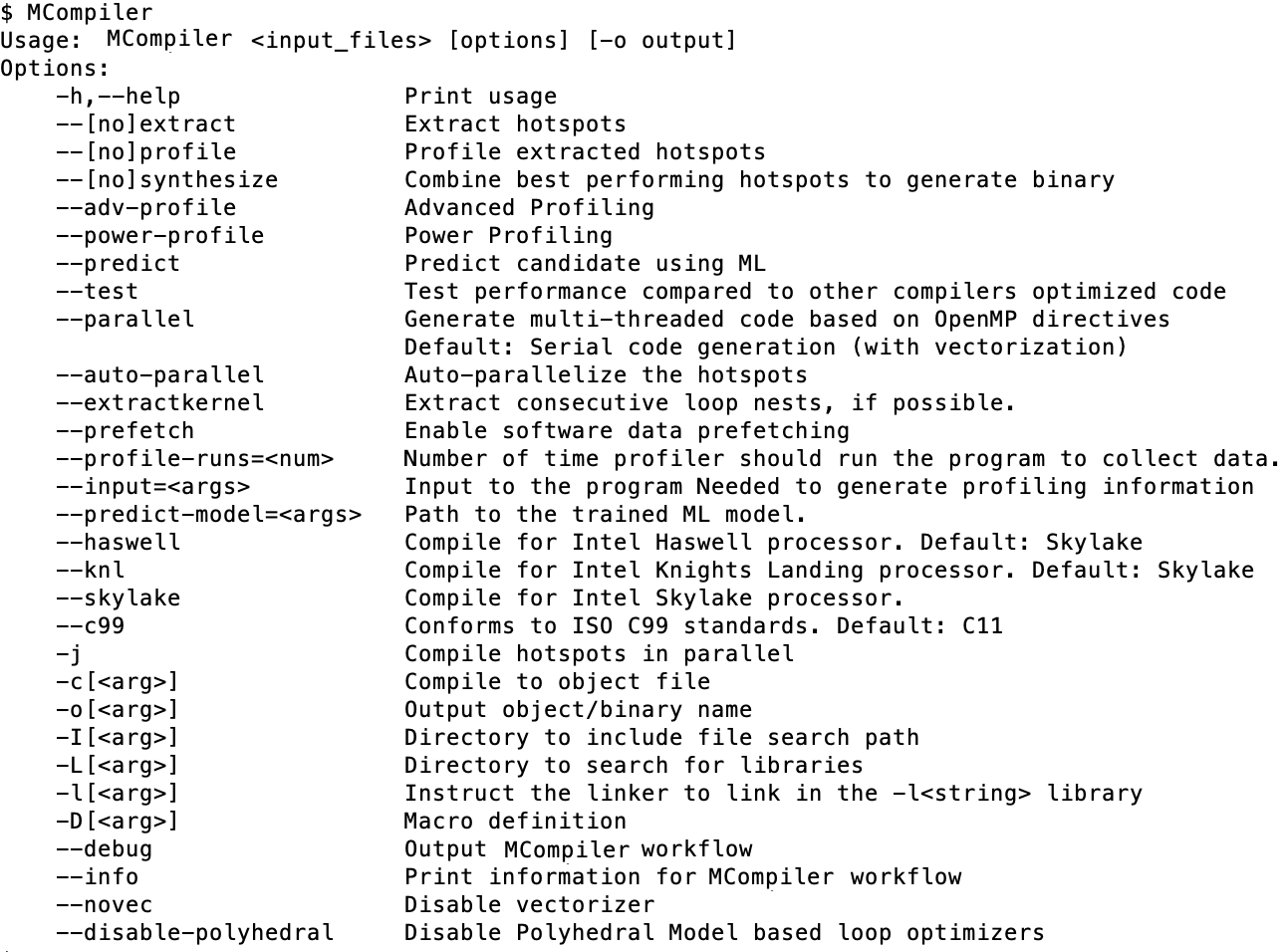}
  \caption{\mC{} command line options}
  \label{fig:mC_options}
\end{figure}
\subsection{Tool for Compiler Researchers}
The \mC{} framework, we believe, is also an important tool for compiler researchers who regularly implement and test their optimization techniques and/or tweak analytical or heuristic models for improving performance for applications.
The framework design also allows for adding new code optimizers and monitoring their performance on entire application or just on particular hotspots.
The command line options for the \mC{} framework are shown in figure \ref{fig:mC_options}.

The framework also allows for training new \ml{} models and using them for making predictions.
Various flags are available for choosing the target architecture and choosing particular optimizations such as auto-parallelization optimizations or enabling particular passes such as data prefetching pass.
The framework also allows for running Advanced Profiler independently, i.e., collect hardware performance counters for all the hotspots in an application while disabling the ML predictions.
\subsection{Instrumenting  Loop Nests to Measure Energy Consumption}
The framework is designed  to be extensible and add other features, in addition to choosing or predicting the most suited code optimizer for loop nests as based on performance.
One such feature very much sought out by developers today is energy consumption analysis and optimization. This sub-section describes the addition of the
energy measurement option for a loop nest.
The \mC{} driver invokes this part of the framework, as shown in figure \ref{fig:power_framework}, if the \texttt{--power-profile} flag is set.
The modifications to the framework required for collecting and reporting  energy measurements are as follows.

First, the Extractor instruments the loop nest body with LIKWID~\cite{Treibig10} APIs.
LIKWID uses the RAPL interface~\cite[Chapter~14.9]{IntelDevManualV3} to measure the consumed energy on the package (socket) and DRAM level.
The Extractor adds \texttt{LIKWID\_MARKER\_INIT} and \texttt{LIKWID\_MARKER\_START(<LOOP ID>)} statements before the loop nest body and adds \texttt{LIKWID\_MARKER\_STOP(<LOOP ID>)} and \texttt{LIKWID\_MARKER\_CLOSE} statements after the loop nest body.
Second, the Optimizer compiles the loop files with an additional macro definition \texttt{-DLIKWID\_PERFMON}.
Third, the Power Profiler generates an executable, each optimized by one of the six code optimizers as mentioned in Table \ref{table:compilerflags}.
Next, the Power Profiler runs the executables with \texttt{likwid-perfctr}
which pins the application to a particular processor and produces the energy measurements.
Finally, the Power Profiler generates the CSV report with energy and power results for each loop nest in the applications.

The goal of adding this feature to the \mC{} framework is to provide a user with more information and insight about the application. 
Furthermore, this feature can be used to generate code that minimizes energy consumption on intended architectures and not just the execution time. Or to optimize the energy-delay product.

\subsection{Expanding the Framework with more Code Optimizers and/or with Optimizer Flag Combinations}
The \mC{} framework allows for addition of more code optimizers so as to give more options for generating the optimized applications.
In addition to that, the framework allows for adding different combinations of compiler flags or code optimizer flags to optimize the applications.
This allows users to explore how different code optimizer flags impact the performance of the applications and use \mC{} framework to generate even better performing executables.
By its design the framework can also include auto-tuning frameworks, such as domain-specific auto-tuner called OpenTuner~\cite{Ansel14}, for optimizing applications.
Exploring different combinations of code optimizer flags is beyond the scope of this work.
In this work, we present results with applications optimized using the most influential or recommended flags combinations, for improving performance, from each code optimizer.

\section{Experimental Analysis} \label{sec:result}
This section  describes the experimental methodology and present the results and their analysis.
\begin{figure*}[t]
  \centering
  \includegraphics[height=8cm,width=0.85\paperwidth]{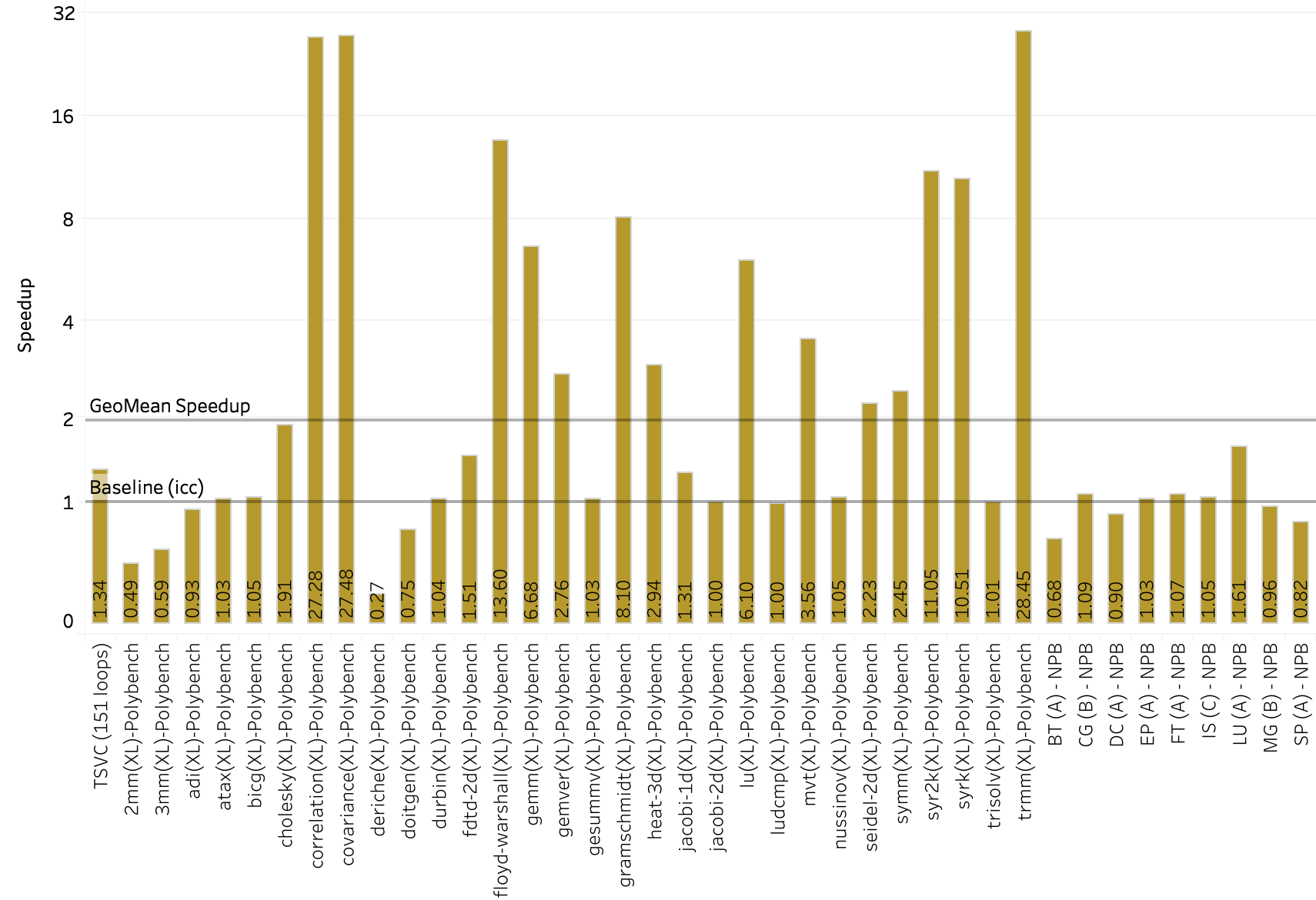}
  \caption{\mC{} Speedup for Serial Benchmarks}
  \label{fig:serial}
\end{figure*}
\subsection{Benchmarks, Code Optimizers and Target Architecture}
Several different benchmark suites are used to evaluate the effectiveness of the \mC{} framework. 
One benchmark suite  used is Test Suite for Vectorizing Compilers (TSVC) by Callahan et al.~\cite{Callahan88} and Maleki et al.~\cite{Maleki11}.
This benchmark was developed to assess the auto-vectorization capabilities of compilers.
Therefore, these loop nests are only used  in the serial code related experiments.
The second benchmark suite used is Polybench~\cite{Polybench}.
This suite consists of 30 benchmarks that perform numerical computations used in various domains, such as linear algebra computations, image processing, physics simulation, etc.
The benchmarks in Polybench have been demonstrated to have performance gain on parallelization, therefore these loop nests are used for auto-parallelized code experiments as well.
The third benchmark suite is NAS Benchmark Suite~\cite{Bailey91}, especially, NPB3.3-SER, NPB3.3-OMP and NPB-ACC~\cite{Xu15}.
These benchmarks are used in serial code, auto-parallelized code and OpenMP parallel code experiments.
Lastly, a set of \texttt{C} benchmarks from SPEC OMP 2012 was used for OpenMP experiments.
The \texttt{train} dataset was used for profiling SPEC benchmarks, whereas the results are shown for \texttt{ref} dataset.
\begin{figure*}[t]
  \centering
  \includegraphics[height=8cm,width=0.85\paperwidth]{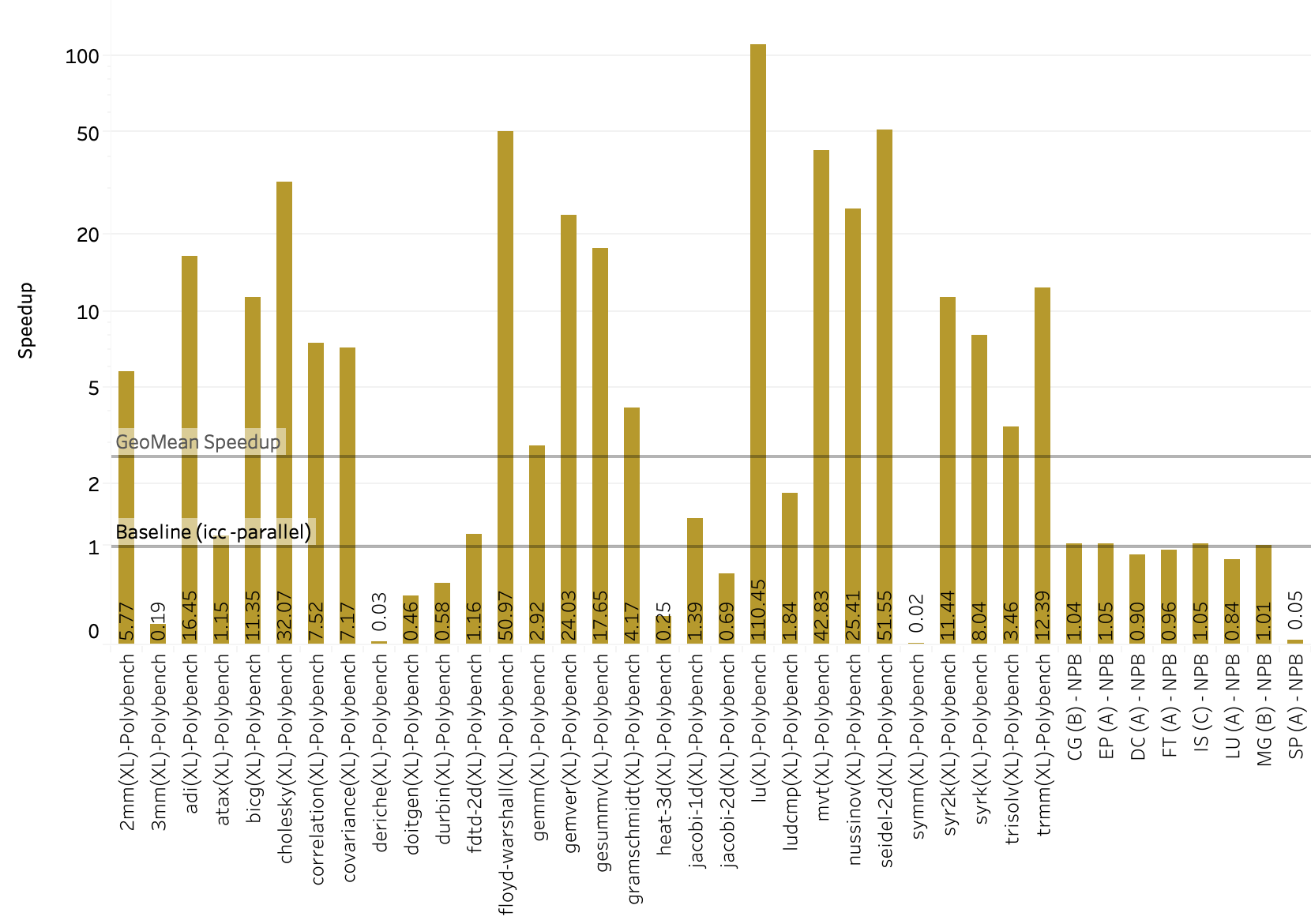}
  \caption{\mC{} Speedup for Auto-Parallelized Benchmarks}
  \label{fig:autopar}
\end{figure*}
Table \ref{table:compilerflags} showed the six code optimizers that have been incorporated in the \mC{} framework.
All six optimizers are used for serial and OpenMP experiments.
Of the six optimizers, only four optimizers (\texttt{icc}, \texttt{pgcc}, \texttt{Polly} and \texttt{Pluto}) can auto-parallelize the serial code and are used for auto-parallelized code experiments.
The baseline for performance comparison  is \texttt{icc} (\texttt{-Ofast -xHost [-parallel]}) compiled benchmarks for all experiments.
\texttt{icc} was chosen as the baseline because \texttt{icc} generated code performed better for more benchmarks than other code optimizers.
The source codes used for the baseline are the original benchmark codes and not the modified source codes generated by the \mC{}'s Loop Extractor.

The target architecture for our experiments is a two-socket, sixteen-core Intel Skylake Xeon Gold 6142. 
Each Xeon processor has 32KB L1 cache, 1MB L2 cache, 22MB L3 cache.
The \sl{} architecture supports SIMD instruction set extensions, i.e., SSE, AVX, AVX2, AVX-512CD and AVX-512F.
Turbo boost is switched off, cores are operating at the maximum frequency, i.e., 2.6 GHz.
For the auto-parallelization and OpenMP experiments, only one thread is mapped per core by setting the environment variables for OpenMP runtimes.
%
\subsection{\mC{} Profiling-Based Search}
This section presents experimental results using the exploratory search by the \mC{} for choosing the most suited code optimizer.
Each application was profiled 3 times for each of the code optimizers and the median execution time was chosen for deciding the most suited code optimizer.
\subsubsection{Serial Code} 
The results are shown in Figure~\ref{fig:serial} with benchmark labels showing the dataset set size in parenthesis and the benchmark suite that a particular benchmark belongs to.
 The GeoMean speedup across the 151 loop nests from TSVC is 1.34x over \texttt{icc}.
The performance  of \mC{} for Polybench benchmarks is usually better than or equal to \texttt{icc} while considering the overheads from \mC{} loop extractions.
As expected, the two polyhedral model based optimizers were chosen as the most suited code optimizer for 55\% of the loop nests from Polybench.
Whereas, for 158 out of 306 (51\%) loop nests from NPB benchmarks, \texttt{icc} is chosen as the most suited code optimizer.
%
\subsubsection{Auto-Parallelized Code} 
These experiments were performed with 32 threads for both profiling the applications and evaluating the performance.
The code optimizers optimized the loop nests with their default setting for statically deciding the profitability of the parallel code and for choosing the runtime settings, such as scheduling policies.

Benchmarks from Polybench, NPB-OMP and NPB-ACC were used in these experiments.
Polybench was shown to have auto-parallelizable loop nests in previous work.
NPB benchmarks use either OpenMP or OpenACC parallel directives and therefore have potential for auto-parallelization. 
The directives were removed from the source code prior to processing by the \mC{}.

The results are shown in Figure~\ref{fig:autopar}. They show that the \mC{} improves performance over \texttt{icc} for 22 of the benchmarks.  Several additional benchmarks have no change in performance.  Five have a significant performance loss, which is explained in Sec.~\ref{sec:analysis} 
\subsubsection{OpenMP Code} 
\begin{figure}[hb]
  \centering
  \includegraphics[height=8cm,width=0.95\columnwidth]{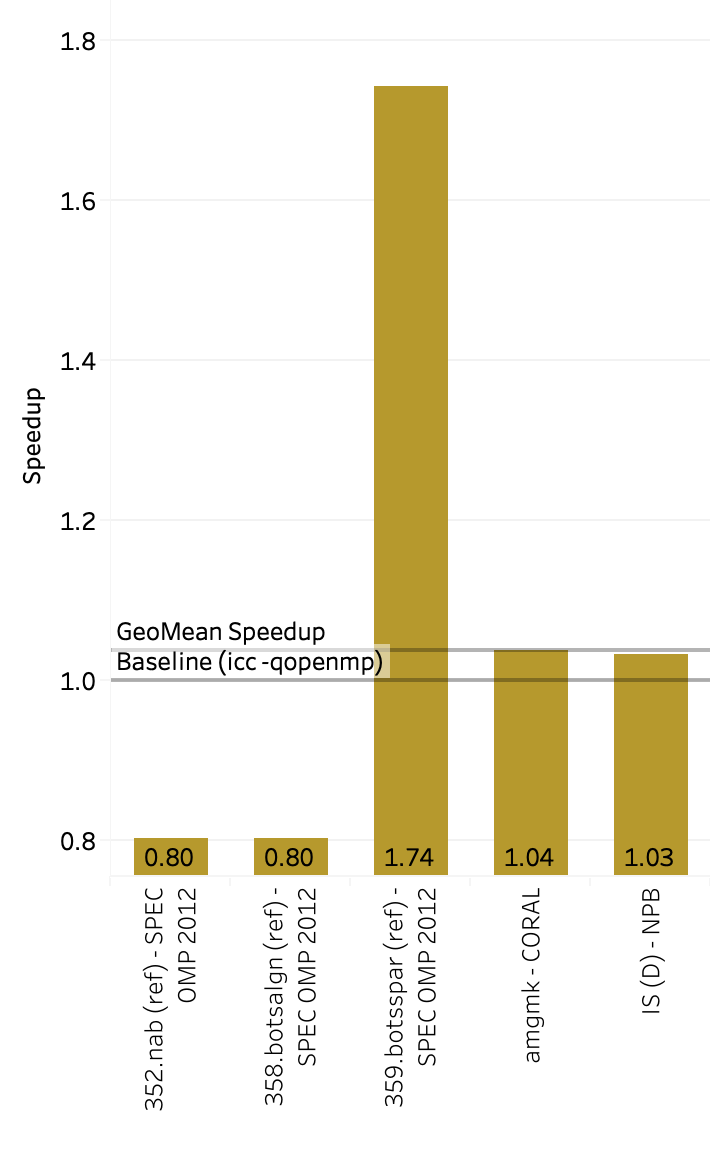}
  \caption{\mC{} Speedup for OpenMP Benchmarks}
  \label{fig:openmp}
\end{figure}

The results are shown in Figure~\ref{fig:openmp}.
Out of a total of 128 loop nests across all benchmarks, \texttt{clang} was chosen as the most suited code optimizer for 28\% loop nests, more than any other code optimizer.
Loop nests that were not marked by OpenMP directives were optimized by the \mC{}  as serial loop nests.
\begin{figure*}[ht]
\captionsetup{justification=centering}
\begin{subfigure}{0.45\textwidth}
  \centering
  \includegraphics[height=8cm]{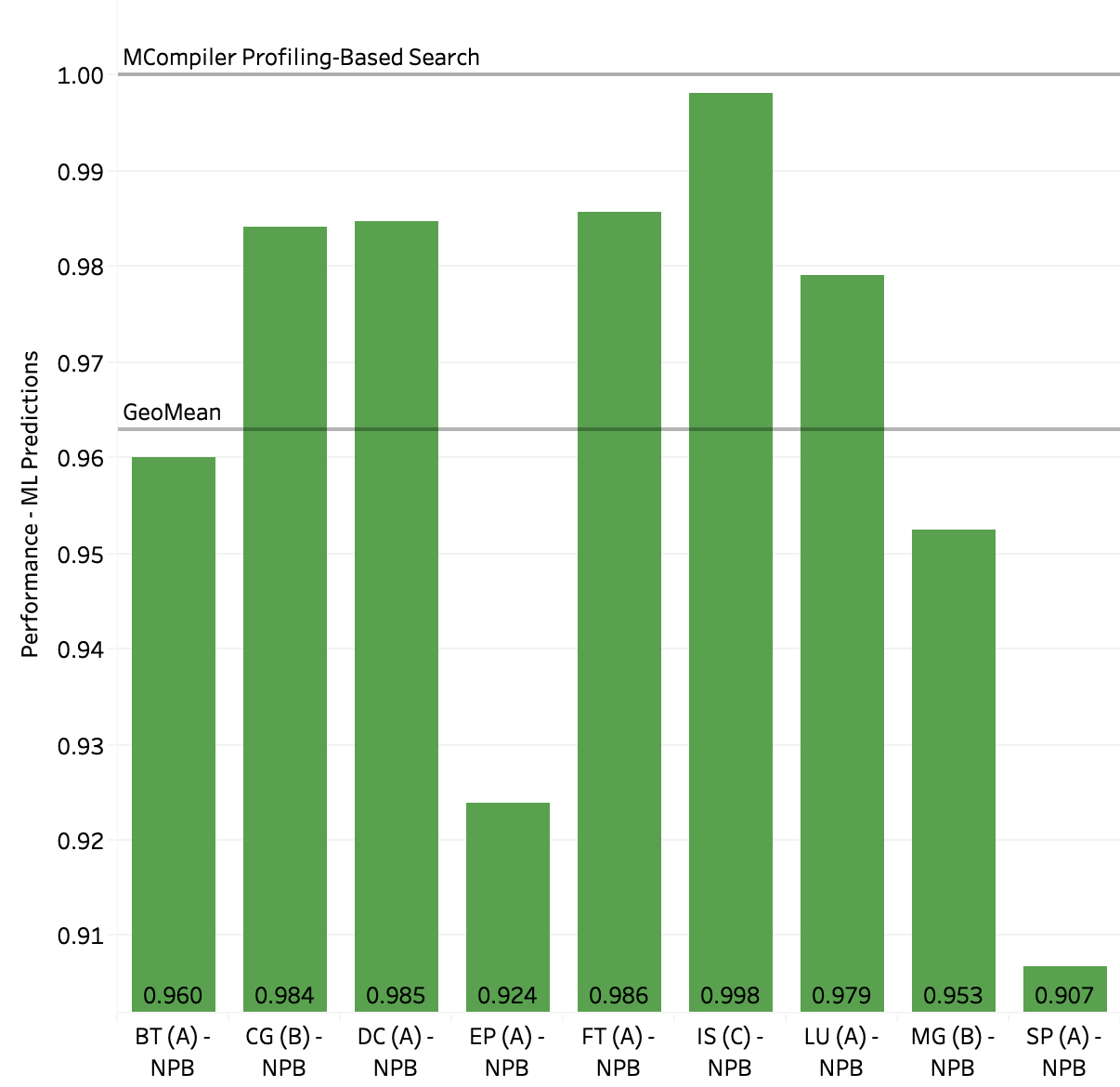}
  \caption{Serial Benchmarks}
  \label{fig:serial-pred}
\end{subfigure}
\begin{subfigure}{0.45\textwidth}
  \centering
  \includegraphics[height=8cm]{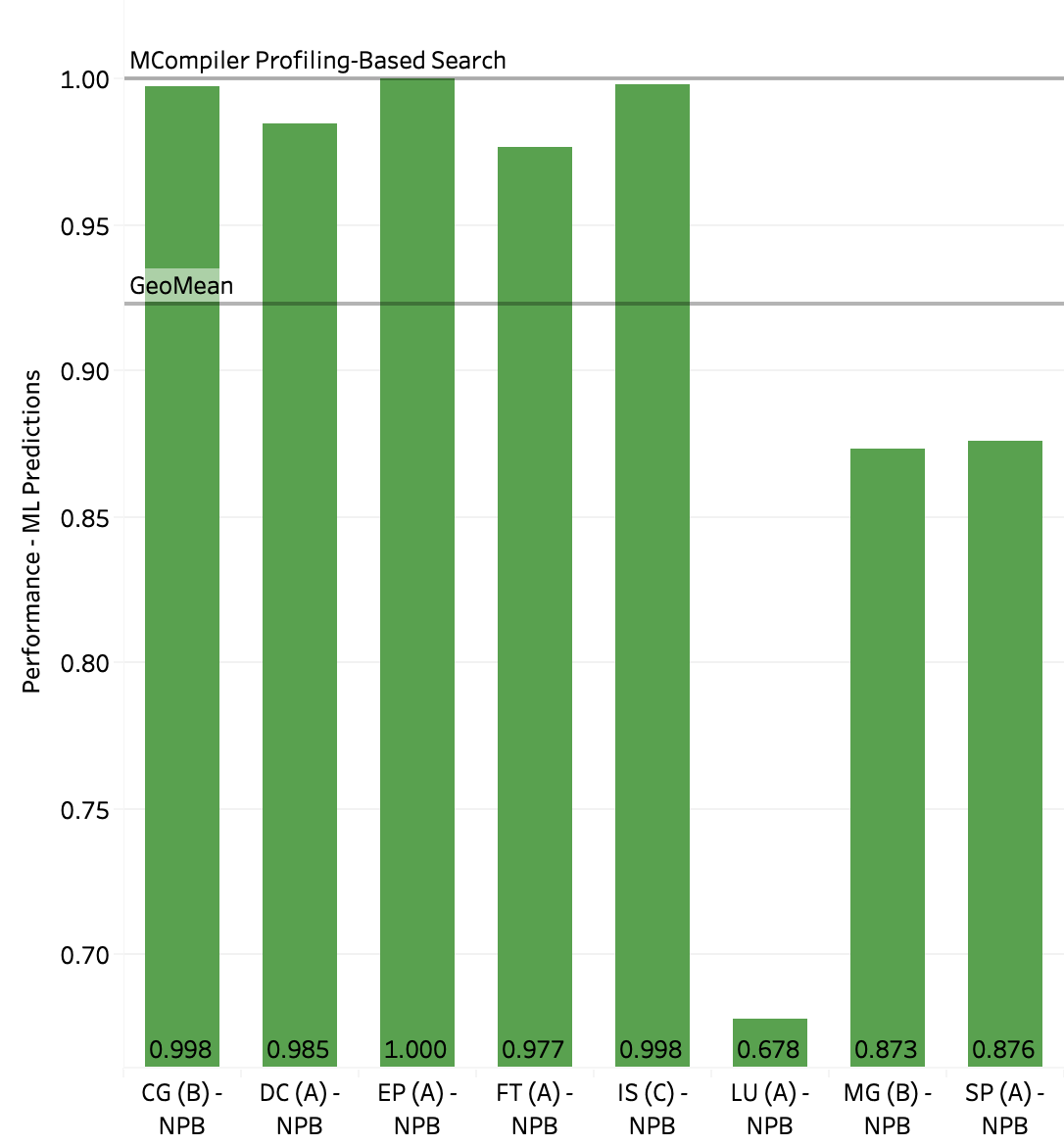}
  \caption{Auto-Parallelized Benchmarks}
  \label{fig:autopar-pred}
\end{subfigure}
\caption{\mC{} + ML Predictions Performance for Serial and Auto-Parallelized Benchmarks}
\end{figure*}
\subsubsection{Analysis of Results}
\label{sec:analysis}
Analysis of the benchmarks that get slowdowns from \mC{}, such as 3mm, deriche (serial), symm (parallel) from Polybench and BT (serial) from NAS benchmark showed that the main reason for performance loss is the extraction of the individual loop nests that contain multiple, consecutive loop nests.
Extraction of an individual loop nest inhibits code optimizers from performing loop optimizations across loop nests.
Such optimizations across loop nests, in general, improve data locality. The impact of improved data locality is even greater for multi-threaded execution.

Another reason for slowdowns can be attributed to the presence of loop nests that have a very short execution time and/or executed multiple times (in a \texttt{while} loop, for example), and performs trivial tasks such as iterating through a linked list.
For such loop nests, the \mC{} extraction adds performance overheads. 

Both of these problems can be solved in the Extractor by adding code analysis to identify consecutive loop nests and, possible statically, identify trivial loop nests with low loop trip count.  This is subject of future work.

Also, the baseline compiler, i.e. \texttt{icc}, analyzes the entire source file and can find more opportunities for optimization, including single-file interprocedural optimizations such as inlining.
For OpenMP benchmarks, we did not expect much performance improvement, since code optimizers lose flexibility to optimize the OpenMP regions due to issues such as early outlining~\cite{Bataev13,Doerfert18} of code.  The one exception seen in Fig.~\ref{fig:openmp} is 359.botsspar, which gets a large speedup with \texttt{Polly}. The reason is the use of \texttt{Polly} optimized code for a loop nest, enclosed in a function, that is called inside a \texttt{omp task} region.
%
%
%
\subsection{\mC{} with \ml{} Prediction}
This section presents experimental results for using the ML predictions for choosing the most suited code optimizer, instead of the profiling-based search in the previous sections.
\subsubsection{\ml{} Model Training and Prediction}
Two ML models were trained, one for predicting the most suited serial, auto-vectorized code optimizer and the other one for the most suited auto-parallelizing code optimizer.
Random Forest (RF) was chosen as our classifier since the accuracy of the RF models for doing multi-class classification was better than other classification algorithms, such as Support Vector Machine (SVM).
The training dataset for training the serial code classifier included loop nests from TSVC and Polybench benchmark suites and has a total of 274 instances (loop nests).
The loop nests from NAS Parallel Benchmarks (NPB) were not included in the training dataset.
Therefore, the experimental results for the \mC{} performance with ML predictions are shown for NPB benchmarks only. 

The auto-parallelized code classifier was trained using the training dataset, which included loop nests from Polybench benchmark suite and has 194 instances (loop nests).
Again, the experimental results for the \mC{} performance with ML prediction are shown for NPB benchmarks only, since these loop nests were never seen by the ML model.
The reason for choosing benchmark suites such as Polybench and TSVC for creating the training dataset was to expose the ML models to a diverse set of loop nests that exhibit different characteristics.
The specifics for creating the training datasets, characteristics of the training dataset and evaluating the models are similar to the work of Shivam et. al.~\cite{Shivam18}.

The properties of the trained RF classifier are as follows.
Maximum depth of the tree was set at 25 after analyzing that the model is neither underfitting nor overfitting on cross-validation.
The maximum sub-categories were set at 15.
The minimum sample count at the leaf node was set at 5.
Lastly, the size of the randomly selected subset of features at each tree node that are used to find the best split is set at 20.

The serial code classifier targets (e.g. most suited code optimizers) were \texttt{clang}, \texttt{gcc}, \texttt{icc} and \texttt{Polly}.
The auto-parallelized code classifier targets were \texttt{icc} and \texttt{Polly}.
\texttt{pgcc} was removed as a target in order to improve the accuracy of the ML models.
In the training dataset, the target for the instances with \texttt{pgcc} as the most suited code optimizer were replaced by the second best code optimizer.
This decision was made after analyzing and tweaking the ML models since the accuracy of the ML models was the priority.  
We left out source-to-source code optimizer, such as Pluto, as a target code optimizer since it requires another compiler to generate code and creates noise for ML models in cases where the performance benefits are not significant from the source-level transformations.

We did not train ML models to predict the most suited code optimizer for the OpenMP loop nests for primarily one reason: the performance of the OpenMP code is largely determined by the presence of OpenMP directives and clauses rather than the properties of the loop nests.
%
\subsubsection{Serial Code} 
The performance results for ML predictions are shown in Figure~\ref{fig:serial-pred} relative to the profiling-based search.
The most predicted code optimizer was \texttt{icc} (51\%), followed by \texttt{clang} (25\%).
The GeoMean performance loss over the profiling-based search is 3.6\%.
The mis-predictions from the ML classifier was found to have a larger impact on performance when most of the execution time is dominated by one or very few kernels, such as in benchmark EP and LU. The effect of a mis-prediction can thus be easily magnified.
\subsubsection{Auto-Parallelized Code} 
The performance results for ML predictions are shown in Figure~\ref{fig:autopar-pred} relative to the profiling-based search.
The most predicted code optimizer was \texttt{polly} (64\%) and the rest was \texttt{icc} (36\%).
The impact of mis-predictions is, in general, higher for auto-parallelized code as compared to serial code.  Still, the GeoMean performance loss over the profiling-based search is rather small - 7.8\%.
\section{Related Work} \label{sec:priorart}
Prior works such as the OptiScope infrastructure presented by Moseley et. al.~\cite{Moseley09} perform function-level and loop-level quantitative comparisons of application compiled by different compilers and/or optimization settings.
Similar to our work, they look at the impact of interaction of optimization techniques for complex target architectures.
But their tool performs binary analysis with the goal of assisting compiler developers in discovering new opportunities and evaluate changes.
The work by Fursin et. al.~\cite{Fursin11} presents an auto-tuning framework that predicts the good combinations of optimizations to improve execution time.
Their tool explores \texttt{gcc} and its optimization flags and uses ML techniques to predict good optimizations based on program features.
Another work, OpenTuner framework by Ansel et. al.~\cite{Ansel14}, searches for the best performing configurations for the domain-specific applications.
Compared to their work, our search space is confined only to the available code optimization candidates and does not require complex heuristics or techniques to find the best performing option.

Prior works that have addressed challenges in compiler optimizations using Machine Learning have focused on auto-vectorization~\cite{Stock12,Watkinson17} and on scalability and scheduling configurations for the parallelism~\cite{Barnes08,Wang09,Tournavitis09}.
Stock et. al.~\cite{Stock12} developed a ML-based performance model to guide SIMD compiler optimizations for vectorizing tensor contraction computations.
However in this work, we explore kernels from a variety of computations to predict an optimizer that can generate an efficient serial code, which includes auto-vectorized code.
Watkinson et. al.~\cite{Watkinson17} in their work use ML models to predict opportunities for auto-vectorization and its profitability across multiple compilers and architectures.
Their work also uses hardware performance counters as ML features and predict opportunities for manual vectorization in loop nests that were not auto-vectorized by the compilers.
In this work, we too explore multiple code optimizers, but we are not concerned with performance gains just related to vector code generation.
We apply ML on a coarser level, in order to predict the most suited code optimizer for serial as well as parallel code.
Tournavitis et. al.\cite{Tournavitis09} use a mix of static and dynamic features to develop a platform-agnostic, profiling-based parallelism detection method for sequential applications. 
Their method requires user's approval for parallelization decisions that cannot be proven conclusively.
They use ML models to judge the profitability on parallelization and to select the scheduling policy.
In contrast, our work uses just the dynamic features to train ML models and we let the ML models choose the most suited candidate that can generate a profitable auto-parallelized code.
In future, we can incorporate the mechanism to predict number of threads and select the scheduling policy as well.

\section{Summary} \label{sec:summary}
This work  presented a compilation framework, called the \mC{}, that optimizes application hotspots for achieving better performance over state-of-the-art compilers.
The framework incorporates optimized loop nest code - serial code, auto-parallelized code or OpenMP code - from a collection of state-of-the-art code optimizers to generate a single executable.
The framework can be used with a profiling-based search to choose the most suited code optimizer for the loop nests.
Experimental results showed that using the \mC{} with a collection of six code optimizers can significantly improve application performance.

The work also showed that one can replace the profiling-based search with an efficient \ml{} based prediction for the most suited code optimizer for each loop nest.
The results show that the \ml{} models can predict the most suited code optimizer with a small performance loss compared to the profiling-based search.
The results also show that the hardware performance counters can capture the inherent characteristics of the loop nests and the \ml{} models based on them make good decisions.

This framework is a tool for compiler researchers to incorporate and analyze the performance of their code optimization techniques and also compare to other code optimizers.

\mC{} framework 
is designed to be extendable with more code optimizers, optimizer flag combinations and more features.

\bibliographystyle{abbrv}
\bibliography{MCompiler_paper}

\end{document}